\documentclass[pra,aps,twocolumn,showpacs]{revtex4}

\usepackage{graphicx}
\usepackage{amsmath,amsfonts,amsbsy,amssymb}
\usepackage{mathrsfs,bbm}

\newcommand{\dual}[1]{\langle #1\rvert}
\newcommand{\ket}[1]{\lvert#1\rangle}

\newcommand{\op}[2]{\ket{#1}\dual{#2}}

\newcommand{\expect}[1]{\langle #1 \rangle}

\begin{document}
\title{Collective uncertainty in partially-polarized and partially-decohered spin-1/2 systems}
\author{Ben Q.\ Baragiola}
\email{quinn.phys@gmail.com}
\author{Bradley A.\ Chase}
\author{JM Geremia}
\email{jgeremia@unm.edu}
\affiliation{Department of Physics and Astronomy, The University of New Mexico, Albuquerque, New Mexico 87131 USA}
\begin{abstract}
It has become common practice to model large spin ensembles as an effective pseudospin with total angular momentum $J=N\times j$, where $j$ is the spin per particle.  Such approaches (at least implicitly) restrict the quantum state of the ensemble to the so-called \textit{symmetric} Hilbert space.   Here, we argue that symmetric states are not generally well-preserved under the type of decoherence typical of experiments involving large clouds of atoms or ions.   In particular, symmetric states are rapidly degraded under models of decoherence that act identically but locally on the different members of the ensemble.  Using an approach [\textit{Phys.~Rev.~A} \textbf{78}, 052101 (2008)] that is not limited to the symmetric Hilbert space, we explore potential pitfalls in the design and interpretation of experiments on spin-squeezing and collective atomic phenomena when the properties of the symmetric states are extended to systems where they do not apply. 
\end{abstract}
\date{\today}
% PACS NUMS
%    03.65.Fd Algebraic methods in quantum mechanics
%    03.65.Yz - Decoherence; open systems; quantum statistical methods
%    34.10.-x General theories and models of atomic and molecular collisions and interactions


\maketitle

\section{Introduction} \label{Section::Introduction}

For a variety of fundamental and technological reasons, there is considerable interest in studying quantum fluctuations in the angular momentum of large atomic/ionic spin ensembles \cite{Kuzmich:2003a,Polzik:2001a,Black:2005a,Chaudhury:2007a,Wineland:1993a,Wineland:1994a,Romalis:2003a,Greiner:2002a,Sadler:2006a,Morrison:2008a,Leibfried2004}.   From a theoretical perspective, modeling such systems is complicated by the fundamental property of quantum mechanics that the Hilbert space $\mathscr{H}_N$ describing $N$ spin-$j$ particles grows exponentially with the number of particles, $\dim\mathscr{H}_N = (2j+1)^N$.  As a result of exponential scaling, it has become common practice to look for dynamical symmetries that reduce the effective dimension of the spin ensemble by restricting its state to a manageable sub-Hilbert space \cite{Stockton:2003a,Chase:2008c}.  One then makes inferences about the properties of the large ensemble based on those of the sub-Hilbert space.  But, of course, the validity of such inferences depends critically on how well the actual spin system respects the symmetries used to formulate the reduced-dimensional description of its quantum state.

Although limited exceptions exist \cite{Bacon:2001a,Chase:2008c}, most work to date on reducing the effective dimension of large spin systems has focussed on the \textit{symmetric group} \cite{Stockton:2003a,Duan:2001a,Duan:2002a}: the sub-Hilbert space $\mathscr{H}_\mathrm{S} \subset \mathscr{H}_N$ spanned by $N$-body states that are invariant under the permutation of particles $\hat{\Pi}_{ij} \ket{\psi} = \ket{\psi}$, $\ket{\psi} \in \mathscr{H}_\mathrm{S}.$  In theory, the symmetric group provides a model of experiments that cannot distinguish between particles during any portion of state preparation, manipulation or measurement.  For spin-1/2 ensembles, the dimension of the \textit{symmetric group} grows only linearly in the number of spin-1/2 particles,  $\dim\mathscr{H}_\mathrm{S} = N+1 \ll 2^N $, making it extremely amenable to simulation and analysis.  Yet, the symmetric states still exhibit interesting multi-particle phenomena, such as entanglement \cite{Stockton:2003a}, spin-squeezing \cite{Kitagawa:1993a} and zero-temperature quantum phase transitions \cite{Morrison:2008a,Emary:2003a,Wang:1973a,Carmichael:1973a,Emary:2003b}.  

This favorable trade-off between manageable size and non-classical behavior has made the symmetric group the sub-Hilbert space of choice for analyzing large spin ensembles--- indeed, any approach that models a large spin ensemble as a collective pseudospin of size $J=N\times j$ \cite{Kuzmich:1998a,Kuzmich:1999a,Kuzmich:2000a,Kuzmich:2003a,Duan:2001a,Duan:2002a,Stockton:2004a,vanHandel:2005a,Bouten:2007b,Morrison:2008a} is grounded at least implicitly in the theoretical underpinnings of particle exchange symmetry \cite{Chase:2008c}.  To justify using the symmetric Hilbert space as a realistic model, two key assumptions are generally made:
\begin{itemize}
\item \textbf{Assumption 1}: The degree of spin polarization that is achieved in practice (such as by optical pumping and possibly additional purification) is sufficient to prepare the ensemble into a state that is well-described by a nearly-pure symmetric state, and ideally by a spin coherent state.

\item \textbf{Assumption 2}: Symmetric states are nearly preserved under low to moderate levels of decoherence, at least of the variety typically encountered in practice, such as that due to spontaneous emission of a far-detuned probe laser.

\end{itemize}
Furthermore, it is generally taken to be true that reasonable laboratory efforts to achieve homogeneous coupling to the electromagnetic fields used to manipulate and measure the ensemble correspond to conditions well-approximated by permutation invariance.  Under these assumptions, several key properties of symmetric collective states, reviewed in Section (\ref{Section::SymmetricStates}), have played a central role in the design and interpretation of experiments involving large spin ensembles:
\begin{itemize}

\item \textbf{Interpretation 1}: Spin-polarized states, such as those obtained by optical pumping, exhibit minimum uncertainty in angular momentum observables transverse to the direction of polarization $\langle \Delta \hat{J}_{\perp_i} \rangle = \sqrt{N/2}$ with respect to the Heisenberg-Robertson inequality (in units where $\hbar=1$)
\begin{equation}
 \langle \Delta \hat{J}_{\perp_1} \rangle \langle \Delta \hat{J}_{\perp_2} \rangle \ge  | \langle  [ \hat{J}_{\perp_1}, \hat{J}_{\perp_2}] \rangle | / 2.
\end{equation} 

\item \textbf{Interpretation 2}: Classical noise, or the uncertainty that results from a classical mixture of spin eigenstates, grows faster than $\sqrt{N}$ and linearly in $N$ for the worst case.  Projection noise scaling that grows faster than $\sqrt{N}$ can be used to diagnose the presence of classical uncertainty in the ensemble.
\end{itemize}
These properties are such fundamental characteristics of the symmetric states that (at least some) research groups have been known to train their members to view a linear increase in spin polarization  coinciding with a square-root increase of spin-projection noise \cite{Wineland:1993a} with atom number as a laboratory signature of a spin coherent state.   These misconceptions have very likely led to the mischaracterization of spin-squeezing in all but perhaps the most recent experiments on spin-noise reduction in large atomic ensembles \cite{SchleierSmith:2009a,Appel:2009a,Windpassinger:2008a,Fernholtz:2008a}.

\subsection{Symmetric versus Collective Decoherence}
\label{Section::SymmetricDecoherence}

For the symmetric Hilbert space $\mathscr{H}_\mathrm{S}$ to remain an accurate description of a spin ensemble's state (provided that the initial state is an element of $\mathscr{H}_\mathrm{S}$),  the system's dynamics must be generated by completely symmetric collective processes:  processes that are themselves permutation invariant and thus expressible in terms of collective operators
\begin{equation} \label{Equation::CollectiveOperator}
	\hat{S} = \sum_n \hat{s}^{(n)}.
\end{equation}
Such operators apply the same single-particle operator $\hat{s} \in \mathfrak{su}(2)$ to each atom in the ensemble, where $ \hat{s}^{(n)} = \hat{1}_1 \otimes \cdots \otimes \hat{1}_{n-1} \otimes \hat{s}_n \otimes \hat{1}_{n+1} \cdots \otimes \cdots \hat{1}_N$ acts non-trivially only on the $n^{th}$ particle.   As such, Eq.\ (\ref{Equation::CollectiveOperator}) is explicitly permutation invariant by construction.

Unfortunately, many of the decoherence models most appropriate for large spin ensembles cannot be described as collective symmetric processes even when the decoherence acts identically on each particle. Consider, the open system dynamics governed by the master equation
\begin{equation} \label{Equation::master}
	\frac{d \hat{\rho}(t)}{dt}  = - \gamma \mathcal{L}[ \hat{s} ]
		\hat{\rho}(t),
\end{equation}
where decoherence acts with the same rate $\gamma$ but locally on every member of the ensemble via the Lindbladian
\begin{equation} \label{Equation::SymmetricDecoherence}
	\mathcal{L}^S[ \hat{s} ] \hat{\rho} \equiv
		\sum_{n=1}^N  \hat{s}^{(n)} \hat{\rho} [\hat{s}^{(n)}]^\dagger  
		- \frac{1}{2}
		\left( [\hat{s}^\dagger \hat{s}]^{(n)} \hat{\rho} +
		\hat{\rho} [\hat{s}^\dagger \hat{s}]^{(n)}  \right) .
\end{equation}
The $(N+1)$-dimensional symmetric-group Hilbert space $\mathscr{H}_\mathrm{S}$ is not preserved under such dynamics, as the Linblad superoperator cannot be expressed in terms of collective operators.  It has thus become common practice \cite{vanHandel:2005a,Stockton:2004a} to study decoherence in spin ensembles by approximating Eq.\ (\ref{Equation::SymmetricDecoherence})
by its associated collective process
 \begin{equation} \label{Equation::CollectiveDecoherence}
	\mathcal{L}^C[ \hat{S} ] \hat{\rho} \equiv  \left[
		\hat{S} \hat{\rho} \hat{S}^\dagger - \frac{1}{2} \left(
			\hat{S}^\dagger \hat{S} \hat{\rho} + \hat{\rho}  \hat{S}^\dagger \hat{S} \right) \right] .
\end{equation}
Eq.\ (\ref{Equation::CollectiveDecoherence}) is more amenable to analysis and simulation because it preserves the $(N+1)$-dimensional symmetric states.  But, it is not always a good physical model.  In atomic systems, for example, a typical source of decoherence comes from spontaneous emission, yet collective radiative processes only occur under stringent conditions such as superradiance from highly confined atoms \cite{Dicke:1954a} and some cavity-QED or spin-grating settings \cite{Black:2005a}.   Even in these cases, the extent to which $N$ atoms behave as a single point-particle dipole moment is imperfect at best. Under typical experimental conditions, where an atomic or ionic ensemble is coupled to a free-space laser probe and the average interatomic spacing is not small compared to the laser wavelength, Eq.\ (\ref{Equation::CollectiveDecoherence}) is just simply the inappropriate model of decoherence \cite{Baragiola:2009a,Chase:2009d}, regardless of how much easier it is to analyze.

\subsection{Main Results}

In this paper, we argue that the statistics of collective angular momentum operators in large spin ensembles are not well-predicted by the behavior of the symmetric group.    Many of our results fall in stark contrast with the conventional wisdom surrounding large spin ensembles:
\begin{itemize}
\item The uncertainty in collective spin obervables for the completely depolarized state of $N$ spin-1/2 particles scales as $\sqrt{N}$ with the number of particles, a scaling that is analytically equivalent to that of a pure spin coherent state.

\item For large ensembles, optical pumping does not produce an approximately pure symmetric state even at high levels of spin polarization.  For example, even with an optical-pumping efficiency of 99.9\%, the purity  of an ensemble with $N \sim 10^6$ spin-1/2 particles is vanishingly small, $\mathrm{tr}[ \hat{\rho}^2] \sim 10^{-409}$, while its overlap with the symmetric group is about $10^{-205}$.  Both decrease exponentially with $N$.

\item For partially polarized ensembles (e.g., incomplete optical pumping), the uncertainty in transverse collective spin observables $\langle \Delta \hat{J}_{\perp i} \rangle$ scales as $\sqrt{N}$ with the number of particles, while the polarization $\langle \hat{J}_\parallel \rangle$ scales linearly in $N$.  Thus, essentially every state of the ensemble corresponding to incomplete optical pumping exhibits the same scaling behavior as an actual spin coherent state.

\item Even if one could prepare a pure initial coherent state, even small levels of decoherence rapidly transform the ensemble state into one that is extremely mixed and very poorly described by a symmetric state.  In fact, we predict that an ensemble with $N\sim 10^5$ particles which has decohered by 20\% (its polarization has dropped to 80\% that of the initial coherent state) has a rather small purity,  approximately $10^{-92,630}$.
\end{itemize}

\section{Collective States of the Ensemble}

Consider an ensemble of $N$ identical spin-1/2 particles described by the single-particle Pauli operators $\hat{\sigma}^{(n)} = ( \hat{\sigma}_\mathrm{x}^{(n)}, \hat{\sigma}_\mathrm{y}^{(n)}, \hat{\sigma}_\mathrm{z}^{(n)})$ and corresponding angular momentum operators $\hat{j}_a^{(n)} =  \hat{\sigma}_a^{(n)}/2$ \footnote{Throughout this paper, the subscripts $a,b=(\mathrm{x},\mathrm{y},\mathrm{z})$ run over Cartesian coordinate labels and $q,r=(+,-,0)$ run over spherical coordinate labels}.  The joint Hilbert space for the entire spin ensemble $\mathscr{H} = \mathscr{H}^{(1)} \otimes \cdots \otimes \mathscr{H}^{(N)}$ has dimension $\dim( \mathscr{H}) = 2^N$, and arbitrary pure states of the ensemble can be expressed in the tensor product basis
\begin{equation}
	\ket{\psi} = \sum_{m_n} c_{m_1,\ldots,m_N} \ket{m_1,m_2,\ldots,m_N}
\end{equation}
where the basis states $\ket{m_1,\ldots,m_N} = \ket{\frac{1}{2},m_1}_1 \otimes \cdots \otimes \ket{\frac{1}{2},m_N}_N$ are simultaneous eigenkets of $[\hat{j}^{(n)}]^2$ and $\hat{j}_\mathrm{z}^{(n)}$: 
\begin{eqnarray}
	[\hat{j}^{(n)}]^2 \ket{m_1,\ldots,m_N} & = &  j_n(j_n+1) \ket{m_1,\ldots,m_N} \\ 
	\hat{j}^{(n)}_\mathrm{z} \ket{m_1,\ldots,m_N} & = & m_n \ket{m_1,\ldots,m_N}.
\end{eqnarray}

Each particle in the ensemble transforms separately under rotation such that $\ket{\psi'} =  [\mathscr{D}^{\frac{1}{2}}(R)]^{\otimes N} \ket{\psi}$,  where $\mathscr{D}^\frac{1}{2}(R)$ is the spin-1/2 rotation operator parameterized by the Euler angles $R=(\alpha,\beta,\gamma)$.  Expressed in the tensor-product basis, the $[\mathscr{D}^{\frac{1}{2}}(R)]^{\otimes N}$ provide a reducible representation for the rotation group but can be decomposed into irreducible components (irreps)
\begin{equation} \label{Equation::IrrepDecomp}
	\mathscr{D}(R) = \bigoplus_{J=J_\mathrm{min}}^{J_\mathrm{max}}  \left[
	 \bigoplus_{i=1}^{d^J_N} \mathscr{D}^{J,i}(R)  \right]
\end{equation}
via the total spin eigenstates
\begin{eqnarray}
	\hat{J}^2 \ket{J,M,i}  & =&  J(J+1) \ket{J,M,i} \\
	\hat{J}_\mathrm{z} \ket{J,M,i}  &=&  M \ket{J,M,i}
\end{eqnarray}
with the collective spin operators $\hat{J}_q = \frac{1}{2} \sum_{n=1}^N \hat{\sigma}_q^{(n)}$ and $J=\mathrm{mod}(N/2,2),\ldots, N/2$.  For each total angular momentum $J$, the quantum number $i=1,\ldots,d^J_N$ distinguishes between the 
\begin{equation}
   	 d^J_N =\frac{N!(2J+1)}{(N/2-J)!(N/2+J+1)!} 
\end{equation}
degenerate irreps with total angular momentum $J$ \cite{Mikhailov:1977a}.    It is readily shown that the degeneracy function satisfies 
\begin{equation} \label{Equation::DNJSum1}
	\sum_{J=0}^{N/2} (2J+1) d_N^J = 2^N.
\end{equation}

\subsection{Generalized Collective States}

In the ``irrep basis,'' arbitrary pure states of the spin ensemble are expressed as
\begin{equation}
	\ket{\psi} = \sum_{J,M,i}  c_{J,M,i} \ket{J,M,i},
\end{equation}
which still requires $2^N$ coefficients [refer to Eq.\ (\ref{Equation::DNJSum1})].  Of course, simply transforming to the irrep basis does not change the effective dimension of the Hilbert space, but it suggests the symmetry that was used to develop the concept of \textit{generalized collective states} in Ref.\ \cite{Chase:2008c}.    Such states are described by the sub-Hilbert space $\mathscr{H}_\mathrm{C} \subset \mathscr{H}_N$ spanned by $N$-particle states that are indistinguishable across the $d_N^J$ degenerate irreps for each total angular momentum $J$.  This generalized permutation symmetry $c_{J,M,i} = c_{J,M,i'}, \forall i,i'$ makes it unnecessary to distinguish basis kets $\ket{J,M,i}$ with respect to their irrep label.  By defining effective basis kets $\ket{J,M}$ on each total-$J$ irrep block, the generalized collective states are
\begin{equation}
	\ket{\psi_\mathrm{C}} = \sum_{J,M} c_{J,M} \ket{J,M},
\end{equation}
with the rescaled coefficients
\begin{equation} \label{Equation::CCoefficients}
		c_{J,M} = \sqrt{\frac{1}{d^J_N}} \sum_{i=1}^{d^J_N} c_{J,M,i},
\end{equation}
where the summation is over the $d_J^N$ copies of the irrep with total angular momentum $J$.  Under this symmetry, $\dim(\mathscr{H}_\mathrm{c}) = (N+2)^2/4$ (for $N$ even) scales only quadratically with the number of particles in the ensemble.   While not as convenient as the linear dimensional scaling of the symmetric group, the $O(N^2)$ scaling of the generalized collective states is still a vast improvement over  exponential scaling and is sufficient to allow simulations with at least a hundred or so particles. 

When studying decoherence and other open-system dynamics of a spin ensemble, it is necessary to work with the density operator of the system, rather than a state vector.  The \textit{collective state density operator} is defined as the direct sum over the reduced density operators $\hat{\rho}_J$ for each total-$J$ irrep block \cite{Chase:2008c}
\begin{equation} \label{Equation::RhoReduced}
	\hat{\rho}_\mathrm{C} \equiv  \bigoplus_J \hat{\rho}_J  =
	 \sum_J \sum_{M,M'} \rho_{J,M;J,M'} \op{J,M}{J,M'} . \,\,
\end{equation}
As defined, the collective density operator restricts against coherence between irrep blocks.  It is shown in Ref.\ \cite{Chase:2008c} and in Section (\ref{Section::Processes}) that symmetric maps exhibit a type of super-selection property, which prevents them from generating coherences between irrep blocks.   Generalized collective states are therefore sufficient to model any dynamics of the form in Eq.\ (\ref{Equation::SymmetricDecoherence}), provided that the initial state satisfies Eq.\ (\ref{Equation::RhoReduced}).

\subsubsection{Irrep Populations and Purity}

The structure of generalized collective states can be analyzed by considering the fraction of the population that resides within each total-$J$ irrep block
\begin{equation}
	p_J = \mathrm{tr}[ \hat{\rho}_J ]
\end{equation}
where $\hat{\rho}_J$ is the reduced density matrix defined in Eq.\ (\ref{Equation::RhoReduced}).  The overlap of a generalized collective state with the symmetric group is therefore given by $p_{N/2} = \mathrm{tr}[ \hat{\rho}_{N/2}]$.  Another important distinction between symmetric and generalized collective states is that the collective states can be mixed over total-$J$ irrep blocks even if all of the reduced density operators $\hat{\rho}_J$ are internally pure.  From Eq.\ (\ref{Equation::CCoefficients}) the purity of the full density operator $\hat{\rho}$ is given by
\begin{equation}
	\text{Purity} = \mathrm{tr}[ \hat{\rho}^2 ] = \sum_J \frac{1}{d_J^N} \mathrm{tr}[ \hat{\rho}_J^2 ] .
\end{equation}

\subsubsection{The Symmetric States} \label{Section::SymmetricStates}

The symmetric states previously considered for large spin ensembles are a special case of the generalized collective states \cite{Chase:2008c}:  $\mathscr{H}_\mathrm{S} \subset \mathscr{H}_\mathrm{C}$ is spanned by the maximal angular momentum manifold: $c_{J,M}=0$ for $J \neq N/2$ and $d^{N/2}_N = 1$.    Since the symmetric states only have support on the maximum-$J$ irrep, their trace vanishes on all irrep blocks except for $J=N/2$, $\mathrm{tr}[ \hat{\rho}_J ] = \delta_{J,N/2}$,
thus providing a simple test to determine whether a collective state is also symmetric.  The symmetric states can be defined equivalently as the manifold of states that can be reached from the maximum $\hat{J}_\mathrm{z}$ eigenstate $\ket{N/2,N/2} \leftrightarrow \ket{\frac{1}{2},\ldots,\frac{1}{2}}$ using only maps generated by collective operators, Eq.\ (\ref{Equation::CollectiveOperator}) (the $z$-polarized state $\ket{N/2,N/2}$ is clearly permutation invariant).   

The \textit{spin coherent states} are a special case of the symmetric collective states, defined by the manifold of states that are simply connected to the $z$-polarized state $\ket{N/2,N/2}$ by a rotation
\begin{equation}
	\ket{\theta,\phi} = \mathscr{D}(\theta,\phi) \ket{N/2,N/2}.
\end{equation}
It is well-known, and readily shown, that the expectation value along the direction of spin polarization $\hat{J}_{\theta,\phi} = \mathscr{D}(\theta,\phi) \hat{J}_\mathrm{z}  \mathscr{D}^\dagger(\theta,\phi)$ is given by
\begin{equation}
	 \dual{\theta,\phi} \hat{J}_{\theta,\phi} \ket{\theta,\phi}
	 	=  \dual{N/2,N/2} \hat{J}_\mathrm{z} \ket{N/2, N/2} = \frac{N}{2}
\end{equation}
with $\langle \Delta \hat{J}_{\theta,\phi} \rangle =0$ while the transverse expectation values vanish for the spin coherent state
\begin{eqnarray}
	\langle \hat{J}_{\perp_1} \rangle & = & \frac{1}{2} \dual{N/2,N/2} (\hat{J}_+ + \hat{J}_-) \ket{N/2,N/2} = 0 \\
	\langle \hat{J}_{\perp_2} \rangle & = & \frac{i}{2} \dual{N/2,N/2} (\hat{J}_+ - \hat{J}_- )\ket{N/2,N/2} = 0, \quad
\end{eqnarray}
but the variances do not
\begin{equation}
	\langle \Delta^2 \hat{J}_{\perp_i} \rangle = \frac{1}{4}  \dual{N/2,N/2} \hat{J}_+ \hat{J}_- \ket{N/2,N/2} - 0 = \frac{N}{4}.
\end{equation}
The scaling of the spin projection noise $\langle \Delta \hat{J}_{\perp_i} \rangle = \sqrt{N}/2$ for a coherent state is the basis for \textbf{Interpretation 1}, described in Sec.\ \ref{Section::Introduction}.

The completely mixed state of a spin system with total angular momentum $j$ is given by $\hat{\rho}=\hat{1}_{2j+1} / (2j+1)$, and therefore the completely mixed symmetric state is 
\begin{equation} \label{Equation::MixedSymmetricState}
	\hat{\rho}^S_\mathrm{mixed}  = \left( \frac{1}{N+1} \right) \hat{1}_{N/2} \oplus \hat{0}_{N/2-1} \oplus \cdots \oplus \hat{0},
\end{equation}
which has the property that it is completely depolarized with respect to all collective spin operators
\begin{equation}
	\langle \hat{J}_a \rangle = \frac{1}{N+1} \mathrm{tr}[ \hat{J}_a] = 0.
\end{equation}
The variance in collective spin observables 
\begin{equation} \label{Equation::DepolarizedUncertaintyCollective}
	\langle \Delta^2 \hat{J}_a \rangle = \frac{1}{N+1} \mathrm{tr}[ \hat{J}_a^2] 
	- \langle \hat{J}_a \rangle^2  = \frac{N(N+2)}{12}
\end{equation}
is a direct consequence of the permutation-invariance constraint $\mathrm{tr}[ \hat{\rho}_{N/2}]=1$.  This linear scaling of the spin projection noise for the mixed state $\hat{\rho}^S_\mathrm{mixed}$ is the basis for \textbf{Interpretation 2}, described in Sec.\ \ref{Section::Introduction}.

\subsubsection{The Completely Mixed Collective State}

When permutation-invariance is lifted (retaining invariance only over the degenerate copies of irreps), the completely mixed state of the $N$ spins generalizes to 
\begin{eqnarray} \label{Equation::MixedGCState}
	\hat{\rho}^C_\mathrm{mixed}
	 & = &  \bigoplus_J  \frac{1}{d_N^J} \bigoplus_{i=1}^{d_N^J} \frac{\hat{1}_J}{2J+1}
		= \frac{1}{2^N} \bigoplus_J  d_N^J \hat{1}_J. 
\end{eqnarray}
That is, for each irrep contribution to the direct sum, the elements of the density operator are given by the ratio of the degeneracy of that irrep to the total dimension of the Hilbert space, 
\begin{eqnarray}
	\hat{\rho}_{J,M;J,M'}  & = & 2^{-N} d_N^J \delta_{M,M'} \\
	& = & \frac{N!(2J+1) \delta_{M,M'}}{2^N(N/2-J)!(N/2+J+1)!}, \nonumber
\end{eqnarray}
precisely as would be expected.  Normalization of the completely depolarized state is readily verified using Eq.\ (\ref{Equation::DNJSum1}).  Once again, the expectation values of all collective angular momentum operators vanish
\begin{equation}
	\langle \hat{J}_a \rangle = 0,
\end{equation}
but their variance does not
\begin{eqnarray}
	\langle \Delta^2\hat{J}_a \rangle & = & \frac{1}{2^N} \mathrm{tr} \left[
		\bigoplus_J d_N^J \hat{J}_a^2 \right]  - \langle \hat{J}_a \rangle^2 \nonumber \\
		& = &  \sum_{J=0}^{N/2} \frac{J(J+1)d_N^J}{3 \cdot 2^N} = \frac{N}{4}.
\end{eqnarray}
In fact, the uncertainty of all collective spin observables with respect to the completely mixed state is quantitatively identical to that of the spin coherent state
\begin{equation} \label{Equation::DepolarizedUncertaintySymmetric}
	\langle \Delta \hat{J}_a \rangle = \sqrt{\frac{N}{2}}.
\end{equation}

\section{Symmetric Dynamics and Collective-State Preserving Processes}
\label{Section::Processes}

The super-operator
\begin{equation} \label{Equation::LCollective}
	\mathcal{L} \hat{\rho}_\mathrm{C} = 
		\sum_{J} \sum_{M,M'} \rho_{J,M;J,M'} f^J_{MM'}
\end{equation}
will preserve collective states $\mathcal{L} : \mathscr{H}_\mathrm{C} \rightarrow \mathscr{H}_\mathrm{C}$ if its action
\begin{equation} \label{Equation::MapAction}
	f^{J}_{MM'} = \mathcal{L} \op{J,M}{J,M'}
\end{equation} can be expressed in the irrep basis in such a way that it does not distinguish between degenerate irreps $f^{J,i}_{MM'}=f^{J,i'}_{MM'}$.  Operators that transform simply with respect to the rotation group
\begin{eqnarray}
	\hat{S} & = &  \bigoplus_J \hat{S}_J = \sum_{J} \sum_{M,M'} s^{J}_{M,M'}\op{J,M}{J,M'} ,
\end{eqnarray}
including the \textit{collective angular momentum operators} $\hat{J}_a$ and all collective operators $\hat{S} = \sum_{n=1}^N \hat{s}^{(n)}$ formed from $\hat{s}^{(n)}\in \mathfrak{su}(2)$, satisfy the requirement of invariance over degenerate irreps by construction.

But as discussed in Section (\ref{Section::SymmetricDecoherence}), processes that are only symmetric over local single-particle super-operators, 
\begin{equation} \label{Equation::SymmetricL}
	\mathcal{L}^S[ \hat{s} ] \hat{\rho}  =  \sum_{n=1}^N \mathcal{L}^{(n)} [ \hat{s}^{(n)} ] \hat{\rho} 
		= \sum_n \hat{s}^{(n)} \hat{\rho} (\hat{s}^{(n)})^\dagger
\end{equation}
do not transform simply under rotations.   Our present work is made possible by results from our previous demonstration that any symmetric local map of the form in Eq.\ (\ref{Equation::SymmetricL}) can be brought into the form of Eq.\ (\ref{Equation::LCollective}) and therefore preserves collective states \cite{Chase:2008c} .  For the $\mathfrak{su}(2)$ operator 
\begin{eqnarray}
	\hat{s} & = & \vec{s} \cdot \hat{\sigma} =  s_0 \hat{1} + s_+ \hat{\sigma}_+ +
		s_- \hat{\sigma}_- + s_z \hat{\sigma}_z
\end{eqnarray}
expressed in the basis $\{ \hat{\sigma}_-, \hat{\sigma}_+, \hat{\sigma}_\mathrm{z}, \hat{1}\}$, the action of Eq.\ (\ref{Equation::SymmetricL}) can be constructed as 
\begin{equation}
	f^{J}_{MM'} = \vec{s} \cdot \mathbf{g}(J,M,M') \cdot \vec{s}^{\,\dagger}
\end{equation}
from the tensor operator
\begin{equation}
	g_{qr} =  \sum_{n=1}^N \hat{\sigma}_{q}^{(n)}\op{J,M}{J,M'} \hat{\sigma}_{r}^{(n)\dagger}.
\end{equation}
The elements of $\mathbf{g}$ can be derived recursively \cite{Chase:2008c} to give
\begin{eqnarray}
g_{qr} \! \! & =   &  
	 \frac{A_q^{J,M} A_r^{J,M'}}{2J} \times    		  
	 	\label{Equation::MainResult} \\
	&&\left( 1+\frac{\alpha^{J+1}_N(2J+1)}{d^J_N(J+1)}\right)
		\op{J,M+q}{J,M'+r}  \nonumber \\
	&+ &  \frac{B_q^{J,M}B_r^{J,M'}  \alpha^J_N}{2 J d^J_N }\op{J-1,M+q}{J-1,M'+r}  
	 \nonumber
		 \\ 
	&+ &  \frac{\alpha^{J+1}_N D_q^{J,M} D_r^{J,M'} }{2 (J+1) d^{J}_N }
		  \op{J+1,M+q}{J+1,M'+r}  \nonumber 
\end{eqnarray}
where the reduced degeneracies are given by $\alpha^J_N = \sum_{J'=J}^{N/2}d^{J'}_N$ and the coefficients are defined as
\begin{eqnarray}
			A_\pm^{J,M} &= & + \sqrt{(J \mp M)(J \pm M+1)}\\
			A_z^{J,M} &= & M \\
			B_\pm^{J,M} &= & \pm \sqrt{(J \mp M)(J \mp M-1)}\\
			B_z^{J,M} &= & \sqrt{(J+M)(J-M)} \\
			D_\pm^{J,M} &= & \mp \sqrt{(J \pm M+1)(J \pm M+2)} \\
			D_z^{J,M} &= &\sqrt{(J+M+1)(J-M+1)} .
\end{eqnarray}
The three terms in Eq.\ (\ref{Equation::MainResult}) arise from two types of processes: (Term 1) transitions that occur between $M$ levels within a single $J$ irrep; and (Terms 2-3) transitions that couple neighboring irreps with $\Delta J = \pm 1$.  It is this coupling between irreps that prevents maps of the form in Eq.\ (\ref{Equation::SymmetricDecoherence}) from preserving symmetric states and that makes collective models of decoherence inadequate for modeling spin ensembles under most laboratory conditions.

\section{Examples}

We have found simulations of large spin systems to be an invaluable tool for studying the properties of symmetric decoherence.  Even though the effective dimension of the generalized collective states grows faster that that of the symmetric group, $O(N^2)$ rather than $O(N)$, it is still possible to run simulations over a sufficient range to make both qualitative and quantitative predictions.  As such, we have performed simulations aimed at addressing the following specific questions:
\begin{enumerate}
 
\item ``Is it possible to prepare a large spin system into a state that is well-approximated by a spin coherent state, and thus a symmetric state?''  Time-evolving the density operator for a spin system under a model of optical pumping enables us to analyze the purity and irrep structure of the system as it is spin-polarized from  an initial mixed state, including what happens for incomplete polarization.
  
 \item ``Do symmetric states remain a good model of spin ensembles subject to limited decoherence?'' Time-evolving the $N$-particle density operator under a symmetric model of spin depolarization enables us to study the relationship between the expectation value and uncertainty of collective angular momentum operators as well as the irrep structure of the state as it decoheres from an initial coherent state. 
 
\item ``Is the practice of approximating symmetric decoherence models with their associated collective processes justified if only expectation values and uncertainties of collective operators are of interest?''  Time-evolving the system under an entangling Hamiltonian and contrasting the effect of the different decoherence models allows us to compare their collective statistics. 

\end{enumerate}

\begin{figure*}[t!]
\begin{center}\includegraphics{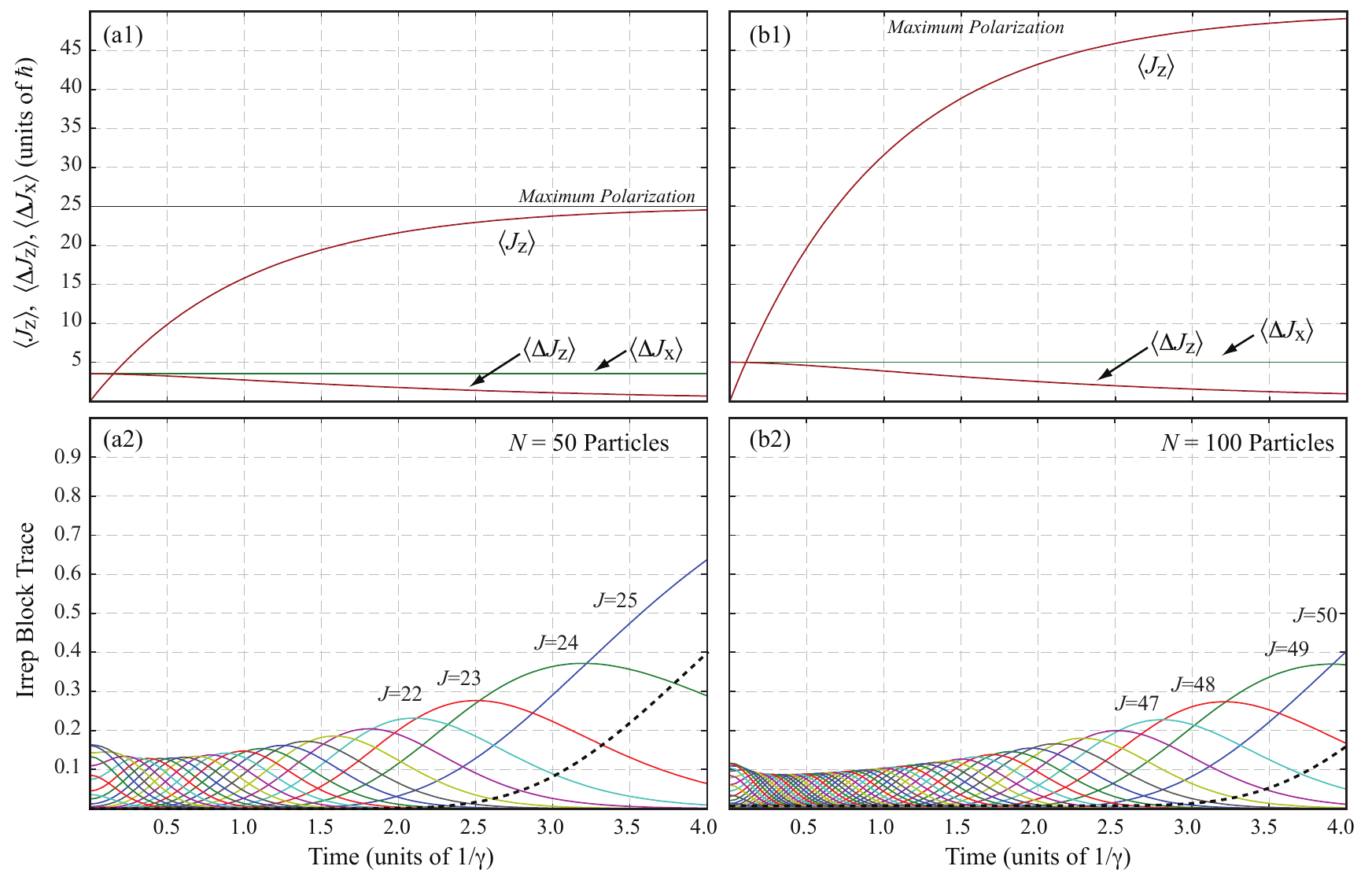}\end{center}
\vspace{-4mm}
\caption{(color online) Simulation of a model of spin-polarization dynamics, corresponding for example to optical pumping computed for $N=50$ (plots a1 and a2) and $N=100$ (plots b1-b2) spin-1/2 particles.  Beginning from the completely-mixed state Eq.\ (\ref{Equation::MixedGCState}), the spin ensemble evolves under the symmetric $z$-axis polarizing channel given by Eq.\ (\ref{Equation::PolarizingChannelSymmetric}).  As the dynamics proceed (a1 and b1), the mean polarization $\langle \hat{J}_\mathrm{z}\rangle$ increases while its uncertainty $\langle \Delta \hat{J}_\mathrm{z}\rangle$ decays.  The transverse uncertainties, $\langle \Delta \hat{J}_\mathrm{x} \rangle = \langle \Delta \hat{J}_\mathrm{y} \rangle = \sqrt{N/2}$ are a constant of the motion.  The evolution of the $J$-irrep block traces (plots a2 and b2) clearly shows that despite $\expect{\hat{J}_\mathrm{z}}$ quickly approaching maximum polarization, the $J < J_{max}$ irrep blocks are still highly populated and the state is quite mixed. \label{figure:figure2}}
\end{figure*}

\begin{figure*}[t!]
\begin{center}\includegraphics{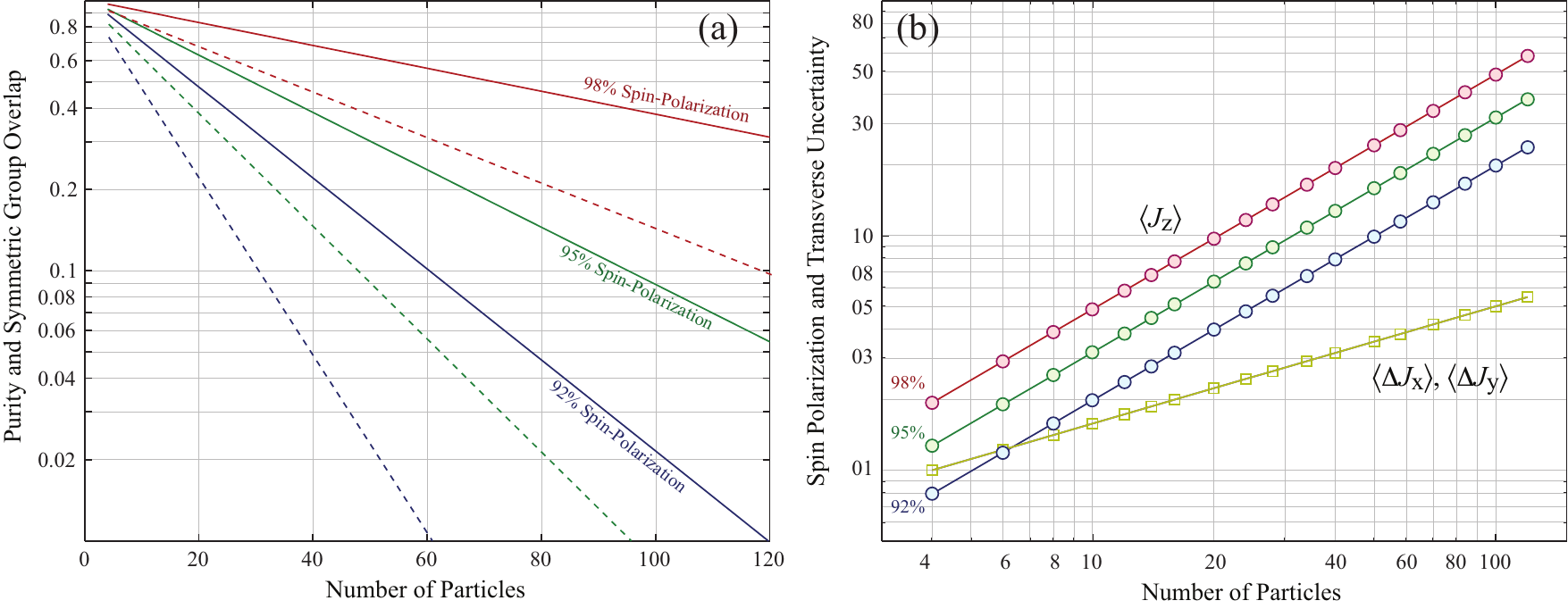}\end{center}
\vspace{-5mm}
\caption{(color online) Demonstration that the symmetric group provides an extremely poor model of a large spin ensemble that is incompletely polarized under the symmetric polarization channel Eq.\ (\ref{Equation::PolarizingChannelSymmetric}) beginning from an initial completely mixed state.   Plot (a) shows the scaling of the purity $\mathrm{tr}[ \hat{\rho}^2]$ and overlap with the symmetric group $\mathrm{tr}[ \hat{\rho}_{N/2}]$ for systems that have been polarized to 92\%, 95\% and 98\% of $\langle \hat{J}_\mathrm{z}^{max} \rangle= N/2$.  Both the purity and overlap with the symmetric group are seen to decrese exponentially in $N$.   Plot (b) shows that the spin-polarization $\langle \hat{J}_\mathrm{z}\rangle$ grows linearly in $N$ while the transverse uncertainties grow as $\sqrt{N}$ even for systems that are incompletely polarized.  \label{figure:figure3}}
\end{figure*}

\subsection{Partial Polarization of the Spin Ensemble}
\label{Section::PartialPolarization}

To determine whether symmetric states, and in particular spin coherent states, provide a good description of a large spin ensemble subject to optical pumping, we considered the symmetric polarizing channel
\begin{equation}  \label{Equation::PolarizingChannelSymmetric}
	\frac{d \hat{\rho}(t)}{dt} = \gamma \, \mathcal{L}^S[\hat{j}_+] \, \hat{\rho}(t) ,
\end{equation}	
which describes the effective spin-1/2 dynamics that arise when radiative excited states are adiabatically eliminated from atoms with two ground states under conditions where the atoms are coupled to a circularly-polarized laser field \cite{Gardiner1991}.  It is readily shown that the steady state corresponding to the symmetric polarizing channel is the spin-coherent state $\ket{\theta=0,\phi=0}$, i.e., the state that is polarized along the positive $z$-axis, with $\langle \hat{J}_\mathrm{z} \rangle = N/2$.  

Under typical laboratory conditions, however, optical pumping does not achieve complete polarization; pumping falls short of reaching the steady state of Eq.\ (\ref{Equation::PolarizingChannelSymmetric}) \cite{Julsgaard:2004a, Fernholtz:2008a}.  Figures \ref{figure:figure2}(a1) and \ref{figure:figure2}(b1) plot the time-evolution of the collective expectation value $\langle \hat{J}_\mathrm{z}\rangle$ and uncertainties, $\langle \Delta \hat{J}_\mathrm{x}\rangle$, $\langle \Delta \hat{J}_\mathrm{y}\rangle$ and $\langle \Delta \hat{J}_\mathrm{z}\rangle$, as the spin ensemble evolves from a completely mixed initial state, Eq.\ (\ref{Equation::MixedGCState}) for $N=50$ and $N=100$ particles.  As expected, the spin polarization $\langle \hat{J}_\mathrm{z} \rangle$ increases monotonically from its initial value of zero, coinciding with a decrease in $\langle \hat{J}_\mathrm{z} \rangle$ as the system progresses toward the maximum-$\hat{J}_\mathrm{z}$ eigenstate.  The transverse uncertainties, $\langle \hat{J}_\mathrm{x} \rangle$ and $\langle \hat{J}_\mathrm{y}\rangle$, are constants of the motion: beginning at $\sqrt{N/2}$ for the completely mixed state and remaining at $\sqrt{N/2}$ at all times as the system progresses toward the $\ket{0,0}$ coherent state.

Figures \ref{figure:figure2}(a2) and \ref{figure:figure2}(b2) show the irrep decomposition and purity of the spin ensemble as it is gradually polarized under the dynamics of Eq.\ (\ref{Equation::PolarizingChannelSymmetric}).  The reduced traces $p_J = \mathrm{tr}[ \hat{\rho}_J ]$ is shown for each of the total-$J$ irrep blocks, for $J=0,1,\ldots,N/2$ (for clarity, only the irrep blocks with $J$ close to $J_{max}=N/2$ are labeled on the plot).  The initial completely mixed state has an extremely small overlap with the symmetric group and a purity that is exponentially small in $N$.  As the dynamics proceed, population is gradually transferred to irreps with increasing angular momentum.  Furthermore, the progression of the state to higher total-$J$ irrep blocks is apparently slower for $N=100$ particles than for $N=50$ particles.  For $N=50$, the maximum-$J$ irrep (symmetric group) begins to show a non-negligible population when the spin polarization is approximately 80\% of $N/2$.  For $N=100$ particles, however, the symmetric group does not begin to be populated until nearly 90\% spin polarization.  The behavior of the purity [dashed lines in Figs.\ \ref{figure:figure2}(a2) and \ref{figure:figure2}(b2)] is more dramatic.  Even at 98\% spin polarization, the state of the ensemble is far from pure: $\mathrm{tr}[ \hat{\rho}^2] < 0.4$ for $N=50$ and $\mathrm{tr}[ \hat{\rho}^2] < 0.2$ for $N=100$.

\begin{figure*}[ht!]
\begin{center}\includegraphics{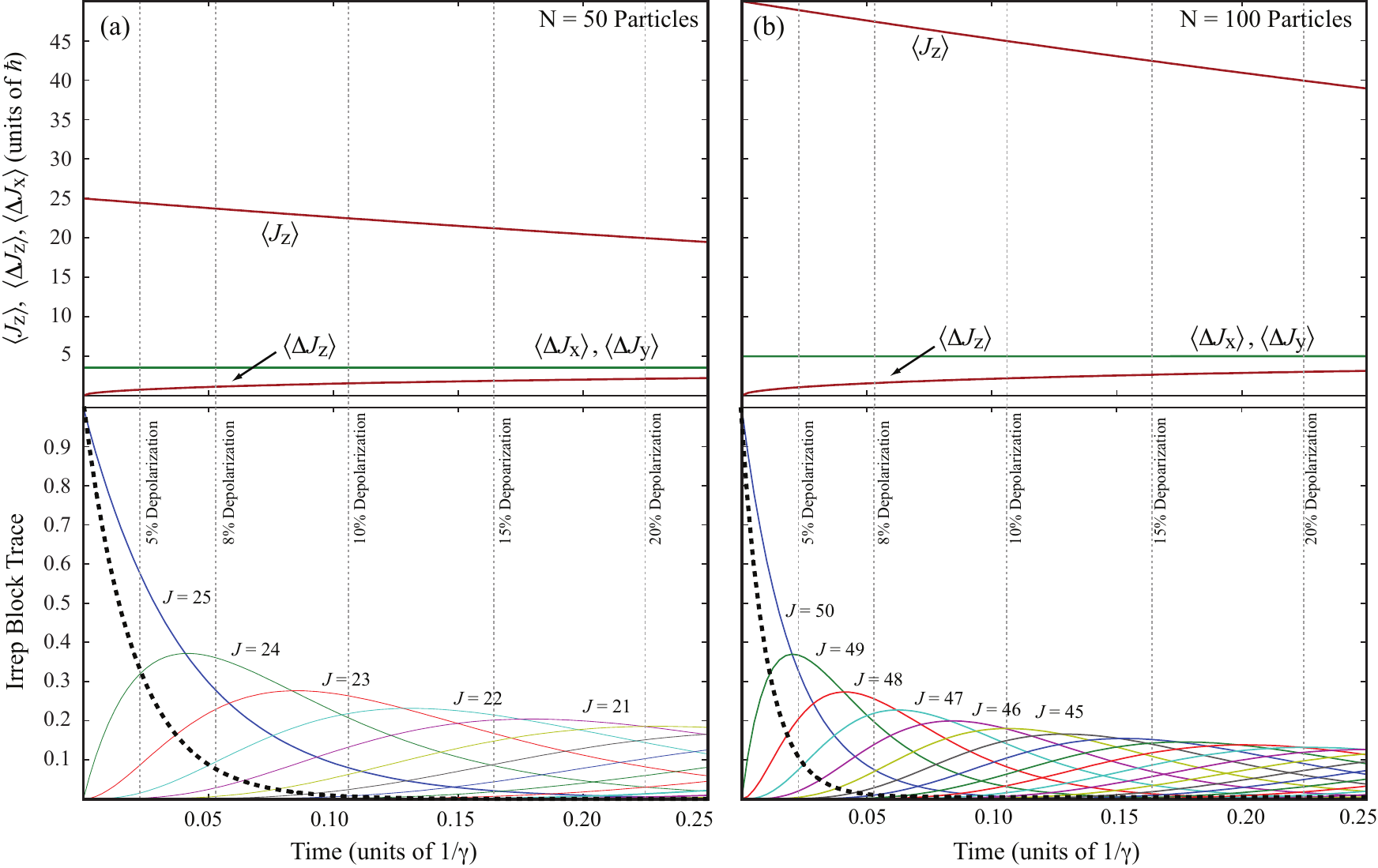}\end{center}
\vspace{-5mm}
\caption{(color online) As the state of a spin ensemble decoheres from a coherent state under the symmetric depolarizing channel Eq.\ (\ref{Equation::DepolarizingChannelSymmetric}), the mean spin-polarization $\langle \hat{J}_\mathrm{z} \rangle$ decreases from its maximum value of $N/2$; however, the transverse uncertainties, $\langle \Delta \hat{J}_\mathrm{x}\rangle$ and $\langle \Delta \hat{J}_\mathrm{y}\rangle$ remain constant with the value $\sqrt{N/2}$.  As the ensemble decoheres, population is transferred out of the symmetric group and into total-$J$ irreps with $J < J_{max}$, as indicated by the evolving irrep traces $\mathrm{tr}[ \hat{\rho}_J(t)]$, as the purity of the state $\mathrm{tr}[\hat{\rho}^2]$ (dotted line) decreases accordingly.  Comparison of the decoherence dynamics for $N=50$ (a) versus $N=100$ (b) particles suggests that the state of the system leaves the symmetric group more rapidly as $N$ increases.
 \label{figure:figure4}}
\end{figure*}

The apparent decrease in purity and overlap with the symmetric group for a given level of spin polarization as the number of particles is increased is explored further in Fig.\ \ref{figure:figure4}.   For each value of $N$, the state of the system is evolved under Eq.\ (\ref{Equation::PolarizingChannelSymmetric}) from a mixed state at $t=0$ until the time when the fractional spin polarization
\begin{equation} \label{Equation::PolarizationFraction}
	f = \frac{\langle \hat{J}_\mathrm{z} \rangle}{N /2}
\end{equation}
achieves a target value.  The corresponding state is then analyzed to determine its overlap with the symmetric group $p_{N/2} = \mathrm{tr}[ \hat{\rho}_{N/2} ]$ and its purity $\mathrm{tr}[ \hat{\rho}^2]$.  Fig.\ \ref{figure:figure4}(a) plots the results for fractional polarizations $f=92\%$, $95\%$, and $98\%$ over the range $4 \le N \le 120$.  As can be seen from Fig.\ \ref{figure:figure4}(a), both the overlaps with the symmetric group (solid lines) and the purities (dotted lines) decrease exponentially with the number of spins $N$ over the range of $N$ that could be analyzed.   Given the consistency of the simulation data as a function of $N$, it seems reasonable to extrapolate the results to higher values of $N$ by fitting the data to an exponential form:
\begin{equation} \label{Equation::NScalingPolarization}
	\mathrm{tr}[\hat{\rho}^2] \approx 10^{-\eta^{op}_p N} \quad\text{and}\quad
	\mathrm{tr}[\hat{\rho}_{N/2}] \approx 10^{-\eta^{op}_s N} .
\end{equation}
Values of the exponents $\eta_p^{op}$ and $\eta_n^{op}$ for various fractional polaizations $f$ are listed in Table \ref{Table::PartialPolarizationExponents}.  The results are quite dramatic, suggesting that even at very high levels of spin polarization, such as $f=99.9\%$, the purity and symmetric overlap achieved by optical pumping in typical experiments are both vanishingly small, eg., $\mathrm{tr}[ \hat{\rho}_{N/2} ] \sim 3 \times 10^{-21}$ for $N=10^5$.  Thus, it seems reasonable to conclude that symmetric states are vastly inadequate for describing such ensembles.

\begin{table}[b]
\caption{Fitted values of the scaling exponents in Eq.\ (\ref{Equation::NScalingPolarization}) for the purity $\mathrm{tr}[\hat{\rho}^2]$ and overlap with the symmetric group $\mathrm{tr}[\hat{\rho}_{N/2}]$ that is partially polarized to a fixed level of spin-polarization $f$ beginning from an initial completely-mixed state. \label{Table::PartialPolarizationExponents}}
\begin{tabular*}{\columnwidth}{@{\hspace{2mm}}c@{\extracolsep{\fill}}cc@{\hspace{2mm}}} \hline\hline
$f$ & $\eta_p^{op}$ & $\eta_s^{op}$ \\ \hline
80.0\% & 0.0856200 & 0.0454300 \\ 
90.0\% & 0.0424700 & 0.0218100 \\ 
95.0\% & 0.0209700 & 0.0106200 \\ 
98.0\% & 0.0084510 & 0.0042460 \\ 
99.0\% & 0.0041670 & 0.0020890 \\ 
99.9\% & 0.0004082 & 0.0002042 \\ 
\hline\hline
\end{tabular*}
\end{table}

Figure \ref{figure:figure3}(b) shows the scaling of the mean polarization $\langle \hat{J}_\mathrm{z}\rangle$ and the transverse uncertainties, $\langle \Delta \hat{J}_\mathrm{x}\rangle$ and $\langle \Delta \hat{J}_\mathrm{y} \rangle$, as a function of $N$ for different levels of optical pumping efficiency.  As expected, the mean polarization scales linearly with $N$.  For incomplete optical pumping, its value is degraded with respect to the maximum value by the factor $f$.  That is, $\langle \hat{J}_\mathrm{z} \rangle = f N/2$.  More surprisingly, perhaps, is that the transverse uncertainties,  $\langle \Delta \hat{J}_\mathrm{x}\rangle$ and $\langle \Delta \hat{J}_\mathrm{y} \rangle$, are always equal to $\sqrt{N/2}$ regardless of the degree of spin polarization.  This result illustrates that there is a fundamental flaw in the laboratory practice of identifying a spin coherent state simply from scaling behavior:  spin polarization that scales as $N$ coinciding with transverse uncertainty that scales as $\sqrt{N}$.  Rather, such an identification is only possible provided with a high-quality, independent measurement of $N$.

\subsection{Decoherence from a Spin Coherent State}

\begin{figure*}
\begin{center}\includegraphics{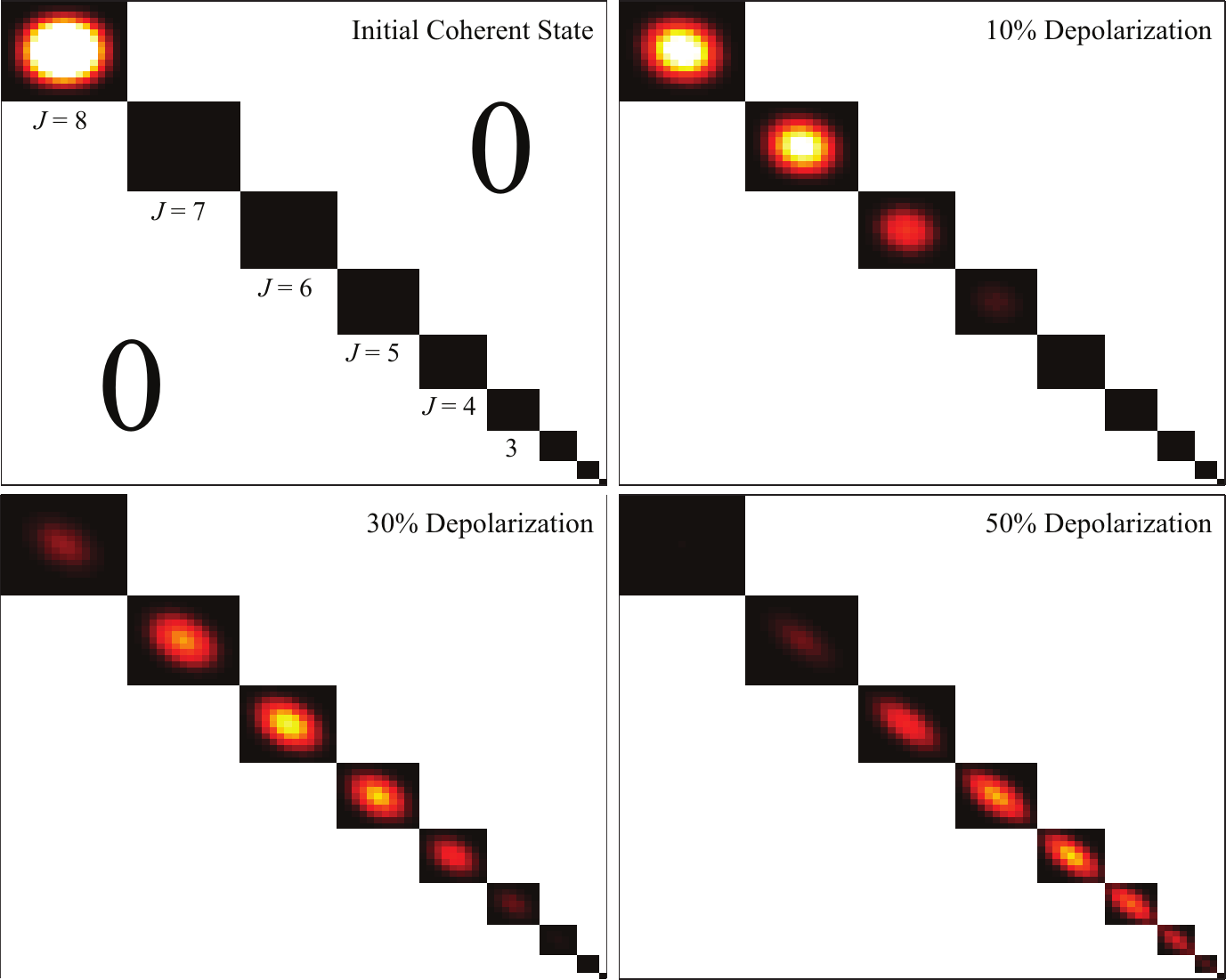}\end{center}
\vspace{-5mm}
\caption{(color online) A fully polarized state undergoing no decoherence is confined to the symmetric space corresponding to the highest $J$-irrep block.  As the state becomes decohered under the depolarization channel Eq. (\ref{Equation::DepolarizingChannelSymmetric}), population from the highest $J$-irrep is transferred to lower $J$-irrep blocks.  Simulations show that even for $N=16$ particles, as seen above, mild amounts of decoherence $\sim 10\%$ significantly deplete the $J=8$ irrep block and the state is driven far from the manifold of symmetric states.  At $30\%$ decoherence, only a vestigial trace population remains in the $J=8$ irrep block.    \label{figure:figure5}}
\end{figure*}

\begin{figure}[hb!]
\vspace{-5mm}
\begin{center}\includegraphics{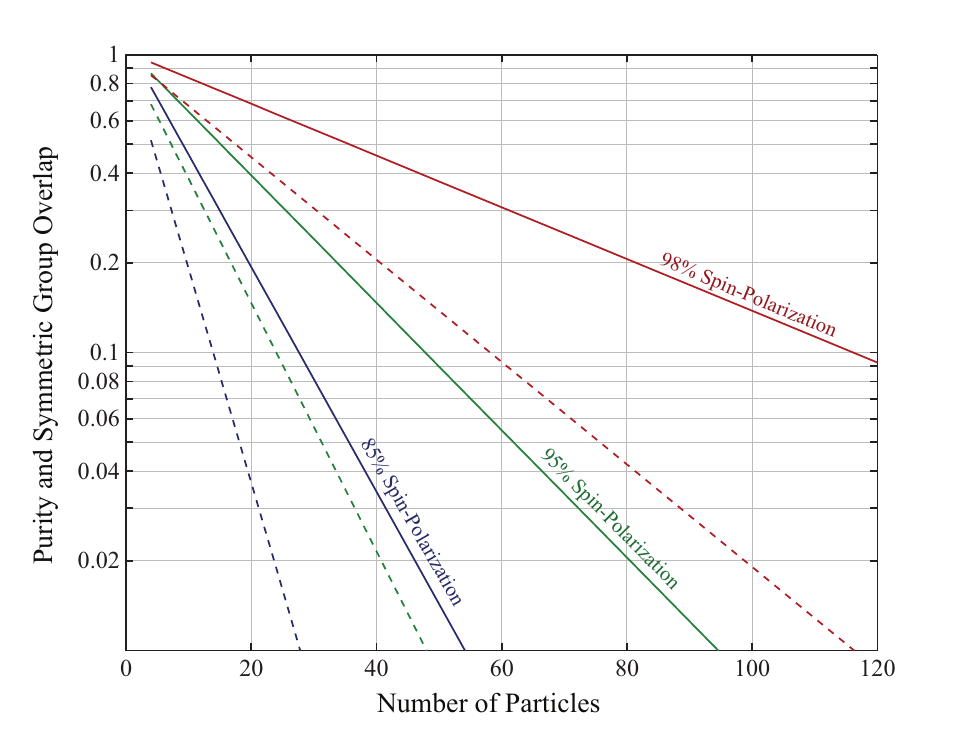}\end{center}
\vspace{-3mm}
\caption{(color online) The purity $\mathrm{tr}[ \hat{\rho}^2]$ (dotted lines) and symmetric group overlap $\mathrm{tr}[ \hat{\rho}_{N/2}]$ (solid lines) are plotted for a system that has decohered  by 2\%, 5\% and 15\% of $\langle \hat{J}_\mathrm{z}^{max} \rangle= N/2$ from an initial spin coherent state under Eq.\ (\ref{Equation::DepolarizingChannelSymmetric}).    \label{figure:figure7}}
\end{figure}

To assess whether symmetric states provide a good model of large spin systems subject to decoherence, we considered the dynamics
\begin{equation}
 \frac{d \hat{\rho}(t)}{d t} = \gamma \, \mathcal{L}^S_{DP} \, \hat{\rho}(t) 
\end{equation}
of an initial spin coherent state subject to the symmetric depolarizing channel
\begin{eqnarray} \label{Equation::DepolarizingChannelSymmetric}
	\mathcal{L}^S_{DP} &=& \left( 
		   \mathcal{L}^S[\hat{j}_\mathrm{x}]  + \mathcal{L}^S[\hat{j}_\mathrm{y}]  
		+ \mathcal{L}^S[\hat{j}_\mathrm{z}]  \right).
\end{eqnarray}
As discussed in the introduction, the symmetric depolarizing channel acts identically but locally on each spin in the ensemble, which is in contrast to the collective analog of Eq.\ (\ref{Equation::DepolarizingChannelSymmetric}), and given by
and collective
\begin{eqnarray} \label{Equation::DepolarizingChannelCollective}
	\mathcal{L}^S_{DP} &=& \left( 
		   \mathcal{L}^C[\hat{J}_\mathrm{x}]  + \mathcal{L}^C[\hat{J}_\mathrm{y}]  
		+ \mathcal{L}^C[\hat{J}_\mathrm{z}]  \right).
\end{eqnarray}

Figure \ref{figure:figure4} plots the time evolution of the expectation value $\expect{\hat{J}_\mathrm{z}}$ and uncertainties, $\expect{\Delta \hat{J}_\mathrm{x} }, \expect{\Delta \hat{J}_\mathrm{y} }$ and $\expect{\Delta \hat{J}_\mathrm{x} }$, of the collective spin operators for ensembles consisting of $N=50$ and $N=100$ particles beginning from the initial $z$-polarized spin coherent state $\ket{\theta=0,\phi=0}$.  As expected, the expectation value $\expect{\hat{J}_\mathrm{z}}$ decreases in time while the uncertainty $\expect{ \Delta \hat{J}_\mathrm{z}}$ increases; however, the uncertainties $\expect{\Delta \hat{J}_\mathrm{y} }$ and $\expect{\Delta \hat{J}_\mathrm{x} }$ are constants of the motion, in contrast to the behavior that would be observed under Eq.\ (\ref{Equation::DepolarizingChannelCollective}).
	
The irrep block traces are plotted as a function of time in the bottom panels of Fig.\ \ref{figure:figure4} for $N=50$ and $N=100$ particles.  It is evident from the plots that the symmetric group quickly becomes a poor description of the state of the spin ensemble:  the trace $\mathrm{tr}[\hat{\rho}_{N/2}]$ of the maximum $J$-irrep block quickly decays.  For $N=50$ particles at 95\% polarization $\mathrm{tr}[\hat{ \rho}_{25} ] < 0.6$, indicating that much of the population has been transferred to lower $J$-irrep blocks.  For $N=100$ at 95\% polarization, the departure from the symmetric group is even more dramatic, with $\mathrm{tr}[\hat{ \rho}_{50} ] < 0.35$.  As $N$ becomes larger, this behavior becomes more pronounced and even minimal decoherence produces significant deviation from the symmetric states.  Analogous to the fitting procedure described in Sec.\ \ref{Section::PartialPolarization}, the purity and symmetric group overlap can be extrapolated to higher numbers of particles according to the exponential fits
\begin{equation} \label{Equation::NScalingDepolarization}
	\mathrm{tr}[\hat{\rho}^2] \approx 10^{-\eta^{dp}_p N} \quad\text{and}\quad
	\mathrm{tr}[\hat{\rho}_{N/2}] \approx 10^{-\eta^{dp}_s N} .
\end{equation}
Values of the exponents $\eta_p^{dp}$ and $\eta_n^{dp}$ for various decoherence levels (measured by the remaining fractional polarization $f$) are listed in Table \ref{Table::DepolarizationExponents}.  Again, the results are dramatic, suggesting that even at low levels of decoherence, the remaining purity and symmetric overlap become exponentially small.  Thus, it also seems reasonable to conclude that symmetric states are vastly inadequate for describing a spin ensemble subject to even small amounts of decoherence even if it were possible to prepare an initial coherent state.  The scalings of the symmetric overlap and purity are plotted as a function of $N$ in Fig.\ \ref{figure:figure7} for various levels of depolarization.

Figure \ref{figure:figure5} depicts the structure of the density operator, expressed in irrep-block basis, for various levels of decoherence.   For the sake of clarity, the figure was generated for a rather small number of atoms $N=16$, however, we have verified that the qualitative results generalize to higher $N$.  As the dynamics proceed from the initial spin coherent state toward the depolarized state, irrep blocks with lower total angular momentum, $J<J_{max}$ become populated.  The steadily decreasing value of the spin polarization $\expect{\hat{J}_\mathrm{z}}$ is therefore a result of two mechanisms: deocherence within each irrep block, and mixing between the blocks.   Throughout this process, the transverse uncertainties of the collective spin operators, $\expect{\Delta \hat{J}_\mathrm{x}}$ and $\expect{\Delta \hat{J}_\mathrm{y}}$ do not increase, even though the uncertainties in individual irrep blocks do.

\begin{table}
\caption{Fitted values of the scaling exponents in Eq.\ (\ref{Equation::NScalingDepolarization}) for the purity $\mathrm{tr}[\hat{\rho}^2]$ and overlap with the symmetric group $\mathrm{tr}[\hat{\rho}_{N/2}]$ that has decohered to a fixed level of spin-polarization $f$ beginning from an initial spin-coherent state. \label{Table::DepolarizationExponents}}
\begin{tabular*}{\columnwidth}{@{\hspace{2mm}}c@{\extracolsep{\fill}}cc@{\hspace{2mm}}} \hline\hline
$f$ & $\eta^{dp}_p$ & $\eta^{dp}_s$ \\ \hline
80.0\% & 0.09263 & 0.04944 \\ 
85.0\% & 0.07182 & 0.03768 \\ 
90.0\% & 0.04946 & 0.02552 \\ 
92.0\% & 0.04165 & 0.02138 \\ 
95.0\% & 0.02552 & 0.01296 \\ 
98.0\% & 0.01719 & 0.008685 \\ 
99.0\% & 0.008685 & 0.004365 \\ \hline\hline
\end{tabular*}
\end{table}

 \begin{figure}[hb!]
 \vspace{0mm}
\begin{center}\includegraphics{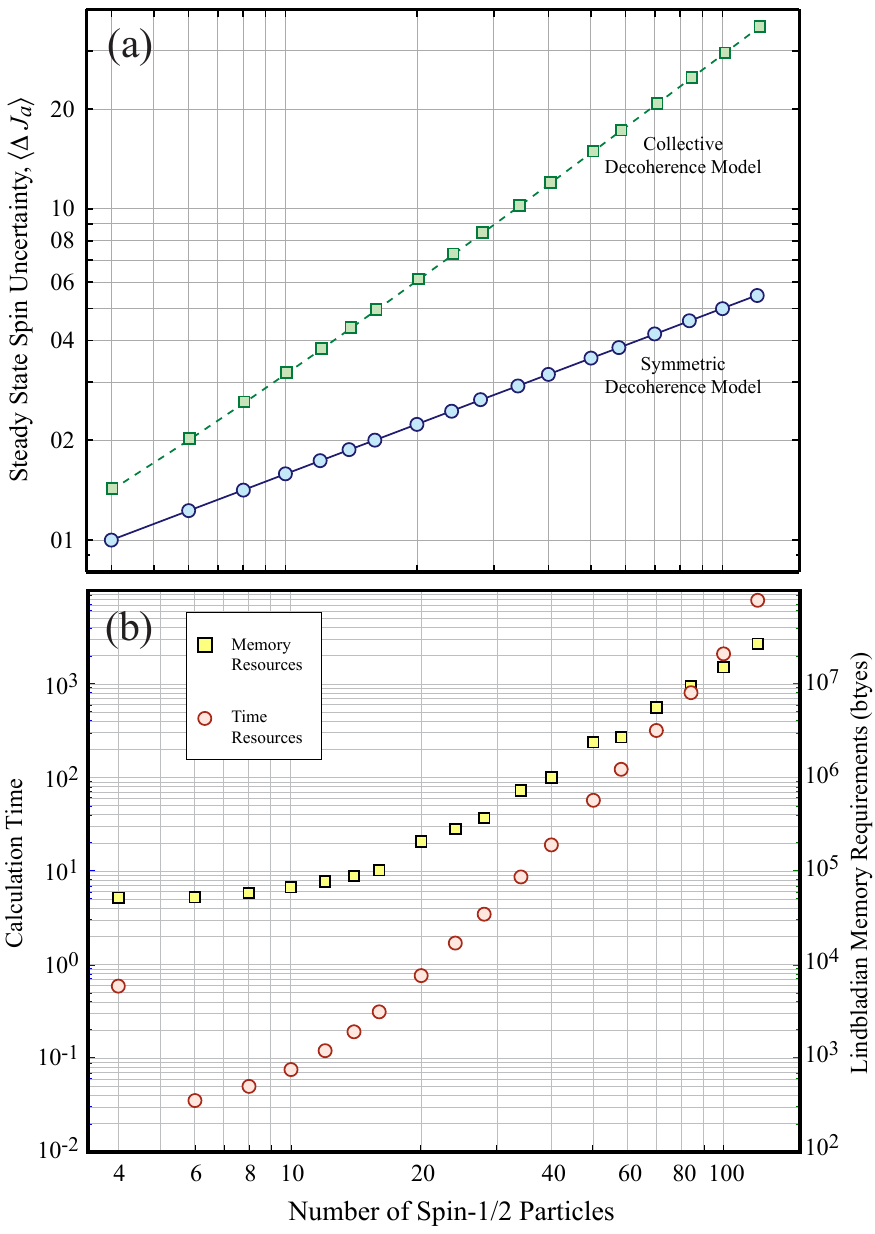}\end{center}
\vspace{-5mm}
\caption{(color online) Plot (a) compares the uncertainty in the collective spin observables $\langle \Delta \hat{J}_q \rangle$ for the steady states of the symmetric depolarizing channel Eq.\ (\ref{Equation::DepolarizingChannelSymmetric}) versus the collective depolarizing channel Eq.\ (\ref{Equation::DepolarizingChannelCollective}) as a function of the number of spin-1/2 particles, $N$.  In the case of collective decoherence, the calculated steady state of $\mathcal{L}^C_{DP}$ agrees with  Eq.\ (\ref{Equation::MixedSymmetricState}) and the calculated uncertainty (squares) matches that of the completely-mixed symmetric state, with $\langle \Delta \hat{J}_a \rangle = \sqrt{N(N+2)/12}$ (dotted line).  For the symmetric decoherence model $\mathcal{L}^S_{DP}$, the calculated steady state agrees with Eq.\ (\ref{Equation::MixedGCState}) and the calculated uncertainty (circles) matches that of the completely-mixed generalized collective state, with $\langle \Delta \hat{J}_a \rangle = \sqrt{N}/2$ (solid line).   In the latter case, the uncertainty scales identically to that of a spin coherent state, despite that it is completely mixed.  Plot (b) indicates the computational resources required to find the steady state as a function of $N$: the memory required to store the Liouville superoperator for $\mathcal{L}^S_{DP}$ (squares, right-side axis) and the time required to find the $\lambda=0$ eigenstate $\hat{\rho}_{ss}$ (circles, left-side axis). \label{Figure::UncertaintyScaling}}
\end{figure}

\begin{figure*}
\begin{center}\includegraphics{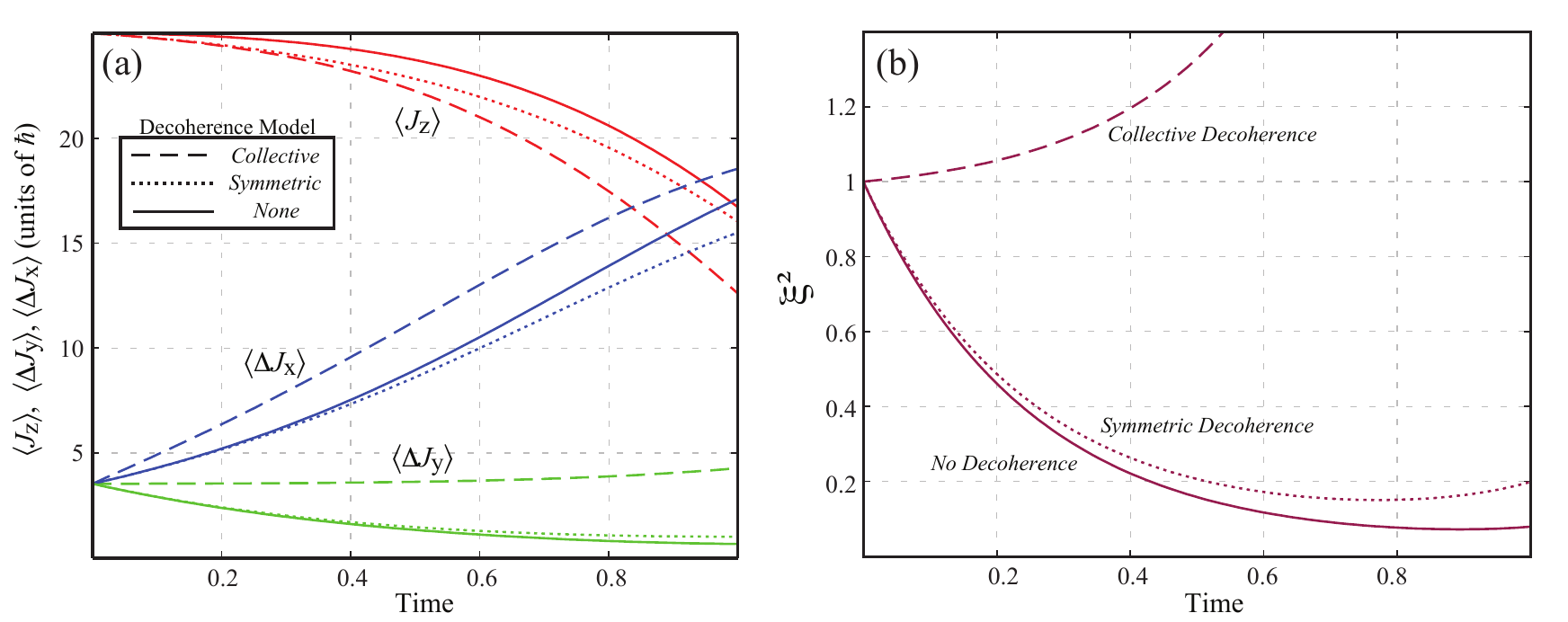}\end{center}
\vspace{-5mm}
\caption{(color online) Simulations of $N=50$ particles evolving under the counter-twisting Hamiltonian $\hat{H} = -i\lambda( \hat{J}_+^2 - \hat{J}_-^2)$ with either collective, symmetric, or no decoherence.  Plot (a) shows that although the expectation $\expect{ \hat{J}_\mathrm{z} }$ and uncertainty $\expect{\Delta \hat{J}_\mathrm{x}}$ evolve in a qualitatively similar manner for the three models, the $\hat{J}_\mathrm{y}$-uncertainty does not:  the symmetric and no decoherence models exhibit a decrease in $\expect{\Delta \hat{J}_\mathrm{y}}$, but the collective model increases.  As a result, the symmetric and no decoherence models exhibit spin-squeezing while the collective model does not, as evidenced in the plot of the squeezing parameter $\xi^2 = N\expect{\Delta^2 \hat{J}_\mathrm{y}}/\expect{\hat{J}_\mathrm{z}}^2$ in plot (b).   \label{figure:figure6}}
\end{figure*}

\subsubsection{Projection Noise of the Depolarized State}

We compared the scaling of the uncertainty in the collective spin observables $\hat{J}_\mathrm{x}$, $\hat{J}_\mathrm{y}$ and $\hat{J}_\mathrm{z}$, for the steady state solutions to the symmetric Eq.\ (\ref{Equation::DepolarizingChannelSymmetric}) versus collective Eq.\ (\ref{Equation::DepolarizingChannelCollective}) mode models of decoherence as a function of the number of particles $N$.  To do so, we determined the steady state density operator $\hat{\rho}_\mathrm{ss}$ corresponding to the decoherece dynamics by solving
\begin{equation}
	 \mathcal{L} \hat{\rho}_\mathrm{ss} = 0 .
\end{equation}
In practice, this is accomplished by expressing both the quantum state $\hat{\rho}$ and the superoperator $\hat{L}$ in their associated Liouville representations, where $\hat{\rho}$ is an $O(N^2 \times 1)$-dimensional column vector and $\hat{L}$ is an $O(N^2 \times N^2)$-dimensional sparse matrix.  The steady state density operator is then given by the eigenvector associated with the $\lambda=0$ eigenvalue of $\mathcal{L}$.     

Our results are illustrated by Fig.\ \ref{Figure::UncertaintyScaling}(a).  The uncertainties of the collective spin observables $\langle \Delta \hat{J}_\mathrm{a} \rangle$ for the steady state of the symmetric  depolarizing decoherence superoperator Eq.\ (\ref{Equation::DepolarizingChannelSymmetric}) as a function of the number of particles $N$ are shown by the circles.  As expected, for a completely depolarized steady state, all of the collective spin uncertainties are identical $\langle \Delta \hat{J}_\mathrm{x} \rangle = \langle \Delta \hat{J}_\mathrm{y} \rangle = \langle \Delta \hat{J}_\mathrm{z} \rangle$.  Furthermore, the uncertainties scale as $\sqrt{N}$ with the number of particles, verified by the agreement of the data points with the predicted uncertainty scaling of the completely depolarized state derived in Eq.\ (\ref{Equation::DepolarizedUncertaintySymmetric}).   For comparison, the uncertainties of the collective spin observables $\langle \Delta \hat{J}_\mathrm{a} \rangle$ were also computed for the steady state solutions to the collective depolarizing superoperator Eq.\ (\ref{Equation::DepolarizingChannelCollective}) as a function of the number of particles [squares in Fig.\ \ref{Figure::UncertaintyScaling}(a)].  Again, all uncertainties are equal for the completely depolarized steady state; however, the scaling with $N$ is linear, in accordance with Eq.\ (\ref{Equation::DepolarizedUncertaintyCollective}).

Figure \ref{Figure::UncertaintyScaling}(b) shows the computational resources required to simulate generalized collective states.  The practical limitation to the maximum value of $N$ that could be analyzed was determined by the required to compute the $\lambda=0$ eigenvector of $\mathcal{L}^S_{DP}$.  Two hours were required to do so for $N=120$ even though the memory required to store the Liouville representation of  Eq.\ (\ref{Equation::DepolarizingChannelSymmetric}), which was only on the order of  30 MB for $N=120$ (although the swap-space consumed by the eigensolver was at least 5 GB for the $N=120$ calculation).

\subsection{Decoherence and Dynamical Spin Squeezing}

As a final example, we compare and contrast the dynamics of a spin system that is subject to Hamiltonian evolution generated by a collective operators as it undergoes symmetric versus collective models of decoherence.  Specifically, we compare evolution under the master equation 
\begin{equation}
	\frac{d \hat{\rho}(t)}{dt} = -i [ \hat{H}, \hat{\rho}(t) ] + \gamma \, \mathcal{L}^S_{DP} \hat{\rho}(t)
\end{equation}
versus the master equation
\begin{equation}
	\frac{d \hat{\rho}(t)}{dt} = -i [ \hat{H}, \hat{\rho}(t) ] +
		\gamma \, \mathcal{L}^C_{DP} \hat{\rho}(t)
\end{equation}
for the ``counter-twisting'' Hamiltonian \cite{Kitagawa:1993a}
\begin{equation}
	\hat{H} = -i \lambda \left( \hat{J}_+^2 - \hat{J}_-^2 \right)
\end{equation}
that has previously been used to study spin squeezing within the symmetric group. 

Dynamics were simulated for a variety of values of $N$, $\lambda$ and $\gamma$ beginning from a $z$-polarized spin coherent state $\ket{\theta=0,\phi=0}$.  Figure \ref{figure:figure6}(a) plots the time-evolution of the collective expectation value $\langle \hat{J}_\mathrm{z} \rangle$ and transverse uncertainties, $\langle \Delta \hat{J}_\mathrm{x} \rangle$ and $\langle \Delta \hat{J}_\mathrm{y} \rangle$, for the specific case of $N =50$, $\lambda = 1/50$ and $\gamma = 4/50$.    As expected, the mean polarization $\langle \hat{J}_\mathrm{z}\rangle$ decreases over time in all three cases.  In the absence of decoherence this apparent depolarization is a byproduct of the increased uncertainty in $\hat{J}_\mathrm{x}$, i.e., the anti-squeezing, even though the state remains pure.  When decoherence is added, increased depolarization is observed, as expected [comparison of the dotted, dashed, and solid lines for $\langle \hat{J}_\mathrm{z} \rangle$ in Fig.\ \ref{figure:figure6}(a)].   As can be seen, this depolarization is noticeably more pronounced under the model of collective decoherence than for the symmetric model.     The time evolution of the transverse uncertainties, $\langle \Delta \hat{J}_\mathrm{x} \rangle$ and $\langle \Delta \hat{J}_\mathrm{y}\rangle$, also shows that the symmetric model of decoherence degrades the collective statistics less so than does collective decoherence.

Each taken alone, neither the uncertainty reduction $\langle \Delta \hat{J}_\mathrm{y} \rangle$ nor the depolarization $\langle \hat{J}_\mathrm{z}\rangle$ assesses the utility of the spin system for precision measurement \cite{Wineland:1993a,Wineland:1994a}.   Figure \ref{figure:figure6}(b) plots the time-evolution of the squeezing parameter
\begin{equation}
	\xi^2 = \frac{N \langle \Delta \hat{J}_\mathrm{y}^2 \rangle}{\langle \hat{J}_\mathrm{z}\rangle^2 + 
		\langle \hat{J}_\mathrm{x} \rangle^2},
\end{equation}
which provides a metric for characterizing the sensitivity of spin-resonance measurements relative to that of a coherent state (for which $\xi^2 = 1$).  It is evident from the plot that, even in the presence of decoherence, the squeezing parameter can drop below $\xi^2=1$.  Even though it would be technically incorrect to refer to the state of the ensemble as a ``spin-squeezed state'' (as such a state as typically defined constitutes a relatively pure symmetric state) \cite{Kitagawa:1993a}, it would appear that a metrological improvement over an actual coherent state is possible even in the presence of  symmetric decoherence under Eq.\ (\ref{Equation::DepolarizingChannelSymmetric}).

\section{Conclusion}

We have identified a number of flaws inherent in using the qualitative properties of symmetric states for modeling the behavior of large spin ensembles.  Even in the most state-of-the-art laboratory settings, experiments involve atom numbers ranging from $5\times10^3 - 10^7$.  Under these conditions, it is  uncommon to find optical pumping efficiencies better than 95\%-98\%, and in many cases the degree of spin polarization may be much worse.  However, even in the best examples of spin polarization, we have found that the resulting state is not well described by a pure spin coherent state even if the expectation values of collective spin operators achieve values that are approximately those of a coherent state.  Furthermore, even when sufficient care is taken in the laboratory to reduce decoherence to minimal levels, highly mixed states with little overlap in the symmetric group are inevitably produced.   As a result of these findings, we conclude that greater care must be exercised when interpreting experiments on large spin systems using scaling laws inferred from the properties of symmetric states.

\begin{acknowledgements}
This work was supported by the NSF under grants PHY-0652877 and PHY-0639994.  Please visit http://qmc.phys.unm.edu/ for more information and to download the code and data files used to generate all figures and data presented in this work. 
\end{acknowledgements}

\vspace{-5mm}
 \bibliography{paper}

\end{document}